\documentclass[lettersize,journal]{IEEEtran}
\usepackage{amsmath,amsfonts}
\usepackage{algorithmic}
\usepackage[T1]{fontenc}  % Use T1 font encoding for extended characters
\usepackage[utf8]{inputenc}  % Use UTF-8 input encoding

\usepackage{algorithm}
\usepackage{array}
\usepackage[caption=false,font=normalsize,labelfont=sf,textfont=sf]{subfig}
\usepackage{textcomp}
\usepackage{stfloats}
\usepackage{url}
\usepackage{verbatim}
\usepackage{graphicx}
\usepackage{cite}
\usepackage{booktabs}
\usepackage{multirow}

\begin{document}

\title{From Audio Deepfake Detection to AI-Generated Music Detection -- A Pathway and Overview}
\author{Anonymous submission}
% \begin{comment}
\author{Yupei Li, Manuel Milling, Lucia Specia, Björn W. Schuller,~\IEEEmembership{Senior Member,~IEEE}
\thanks{Yupei Li, GLAM team, Imperial College London, London, UK, email: yupei.li22@imperial.ac.uk, contact author}
\thanks{Manuel Milling, CHI -- Chair of Health Informatics, Technical University of Munich, Munich, Germany, email: manuel.milling@tum.de}
\thanks{Lucia Specia, GLAM and Llama team, Imperial College London, London, UK, email: l.specia@imperial.ac.uk}
\thanks{Björn W. Schuller, GLAM / CHI -- Chair of Health Informatics, Imperial College London / Technical University of Munich, London, UK / Munich, Germany, email: schuller@tum.de}
\thanks{This research was partially supported by the Munich Center for Machine Learning and the Munich Data Science Institute.}
}
% \end{comment}
%\thanks{Manuscript received April 19, 2021;}}

\markboth{Journal of \LaTeX\ Class Files}%
{Shell \MakeLowercase{\textit{et al.}}: From Audio Deepfake Detection to AI-Generated Music Detection -- A Pathway and Overview}

\IEEEpubid{
  \parbox{\textwidth}{ % Ensure it fits across the entire double-column width
    \centering % Centers the content inside the parbox
    ~\copyright~2024 The Authors. This work is licensed under a Creative Commons Attribution-NonCommercial-NoDerivatives 4.0 License.\\
    For more information, see \url{https://creativecommons.org/licenses/by-nc-nd/4.0/}
  }
}


\maketitle
\begin{abstract}

As Artificial Intelligence (AI) technologies continue to evolve, their use in generating realistic, contextually appropriate content has expanded into various domains. Music, an art form and medium for entertainment deeply rooted in human culture, is seeing an increased involvement of AI into its production. However, the unregulated use of AI music generation (AIGM) tools raises concerns about potential negative impacts on the music industry, copyright, and artistic integrity, underscoring the importance of effective AIGM detection. This paper provides a systematic overview of existing AIGM detection methods. We first establish a four-level detection taxonomy: signal-level, feature-level, watermark, and semantic consistency, organising methods according to the type of trace they exploit. Drawing on the more mature field of audio deepfake detection, we then present a stratified transferability analysis that examines which components may or may not transfer to AIGM detection, and under what conditions. A multi-dimensional classification further organises representative methods along input modality, detection granularity, feature type, model type, detection target, robustness setting, and interpretability. We conclude by discussing implications and proposing directions for future research to address ongoing challenges in the field.

\end{abstract}


\section{Introduction}
\IEEEPARstart{A}{s} multimodal large language models (MLLMs) like GPT-4 \cite{openai2023gpt4} and Whisper \cite{radford2022whisper} continue to evolve rapidly and are increasingly adopted by the public, artificial intelligence-generated content (AIGC) has gained considerable popularity. Among these advancements, audio deepfake has become particularly widespread, creating challenges in the form of potential spam through voice conversion (VC) that mimics intimate voices, as well as public misinformation via text-to-speech (TTS) models that can impersonate political figures.

AI-generated music (AIGM), which has overlap with audio deepfake generation, has seen rapid adoption with platforms like OpenAI’s MuseNet\footnote{\url{https://openai.com/research/musenet}}, Jukedeck\footnote{\url{https://jukedeck.com}}, and AIVA\footnote{\url{https://www.aiva.ai}}. This technology carries profound implications for the music industry, sparking debates around originality, copyright, and artistic value \cite{henry2024impacts} within the music community. For musicians, AI's rapid music production capabilities risk overshadowing artist-created works, which demand significant time, effort, and personal experience \cite{mccormack2019autonomy}. Moreover, the pervasive influence of AI-generated patterns may diminish originality, as artists might unconsciously replicate repetitive AI-driven motifs, hindering innovation \cite{mycka2025artificial}. Additionally, the prevalence of music plagiarism has led to numerous legal disputes, highlighting the need for standardised tools for AIGM detection \cite{malandrino2022adaptive}. For audiences, there is concern that the aesthetic landscape of music could shift in favour of styles driven by AI, potentially altering public taste \cite{torres2025language}. Given the subjective nature of musical judgment, there is a collective reluctance to let AIGM, which lacks authentic emotional depth and creativity, replace human artistry. Therefore, robust AIGM detection is essential to monitor and manage the influence of AI music production, safeguarding the creative integrity of the music community and preventing potential disruptions to both creators and consumers \cite{greyson2025regulatory}.

\IEEEpubidadjcol

Among the efforts to detect AI-generated audio, the majority of attention has been directed towards audio and multimodal deepfake detection. Existing audio deepfake surveys offer fragmented perspectives: Mubarak et al.~\cite{mubarak2023survey} focus on model performance under ASVspoof 2021~\cite{liu2023asvspoof} and ADD 2023~\cite{yi2023add} with limited methodological comparison; Masood et al.~\cite{masood2023deepfakes} and Lin et al.~\cite{lin2024detecting} treat audio as one modality among text, image, and video, limiting depth on audio-specific techniques; Dixit et al.~\cite{dixit2023review} propose a high-level pipeline over a narrow set of approaches; Patel~\cite{patel2023deepfake} examines generation mechanisms without comprehensive detection analysis; and Almutairi and Elgibreen~\cite{almutairi2022review} provide a structured review but only up to 2022. Meanwhile, multimodal deepfake surveys~\cite{xu2025advancements,cao2025survey} address AI-generated content broadly without focusing on AIGM, and music-domain surveys~\cite{mitra2025music,wei2025tools} concentrate on generation techniques rather than detection. As summarised in Table~\ref{tab:survey_comparison}, to the best of our knowledge, no existing survey explicitly focuses on the detection of AIGM, highlighting a significant gap in the current literature.

Although AIGM detection shares its binary classification structure with audio deepfake detection, the two tasks differ along several fundamental dimensions that warrant a dedicated treatment. First, music is an art form governed by musicological constraints such as melodic contour, harmonic progressions, cadential resolution, and long-range form~\cite{hernandez2022music}, which have no direct speech counterpart and require domain-specific features to exploit. Second, speech signals carry prosodic cues (formant trajectories, phonotactics, speaking rate) tied to a small phoneme inventory, whereas music signals span dense harmonic overtones, polyphonic overlap, and a wide dynamic range, making the statistical patterns that signal-level detectors rely upon fundamentally different. Third, music often integrates lyrics with poetic qualities~\cite{Irvine2015}, introducing a multimodal dimension that is absent from most speech-deepfake tasks: inconsistencies between AI-generated melody and lyrics constitute a detection signal that single-modality approaches cannot access. These structural, musicological, and signal-level distinctions mean that methods and representations developed for audio deepfake detection cannot be directly transferred to AIGM, motivating the domain-specific taxonomy and transferability analysis presented in this work.

\begin{table*}[htbp]
\centering
\caption{Comparison of existing surveys related to audio deepfake detection and AI in music.}
\label{tab:survey_comparison}
\begin{tabular}{|p{3cm}|p{1.5cm}|p{2.5cm}|p{3.5cm}|p{3cm}|}
\hline
\textbf{Survey} & \textbf{Year} & \textbf{Main Focus} & \textbf{Coverage of Audio Deepfake Detection} & \textbf{Limitations} \\ 
\hline
Almutairi and Elgibreen \cite{almutairi2022review} & 2022 & Audio deepfake detection overview & General techniques and datasets & Limited to studies before 2022; lacks discussion of recent methods \\ 
\hline
Mubarak et al. \cite{mubarak2023survey} & 2023 & audio deepfake detection models & Evaluation of models using ASVspoof 2021 and ADD 2023 & Dataset scope is limited; lacks broader methodological comparison \\ 
\hline
Masood et al. \cite{masood2023deepfakes} & 2023 & Multimodal deepfake detection & Audio considered as one modality among others & Focuses on multimodal approaches; limited analysis of audio-specific techniques \\ 
\hline
Lin et al. \cite{lin2024detecting} & 2024 & Multimodal deepfake detection & Includes audio alongside image and video modalities & Audio detection methods are not examined in depth \\ 
\hline
Dixit et al. \cite{dixit2023review} & 2023 & Detection pipeline for audio deepfakes & High-level detection workflow & Covers a relatively narrow range of techniques \\ 
\hline
Patel \cite{patel2023deepfake} & 2023 & Deepfake generation and case studies & Brief overview of detection methods & Lacks a comprehensive analysis of detection strategies \\ 
\hline
Mitra et al. \cite{mitra2025music} & 2025 & AI techniques in music generation & Not focused on detection & Primarily reviews music generation rather than detection \\ 
\hline
Wei et al. \cite{wei2025tools} & 2025 & AI tools for music production & Not focused on detection & Focuses on generation technologies \\ 
\hline
\textbf{This work} & \textbf{2026} & \textbf{AIGM detection survey} & \textbf{Comprehensive analysis of detection methodologies for AIGM} & \textbf{First survey dedicated to AIGM detection} \\ 
\hline
\end{tabular}
\end{table*}

\begin{figure*}[ht!]
\centering
\includegraphics[width=\textwidth]{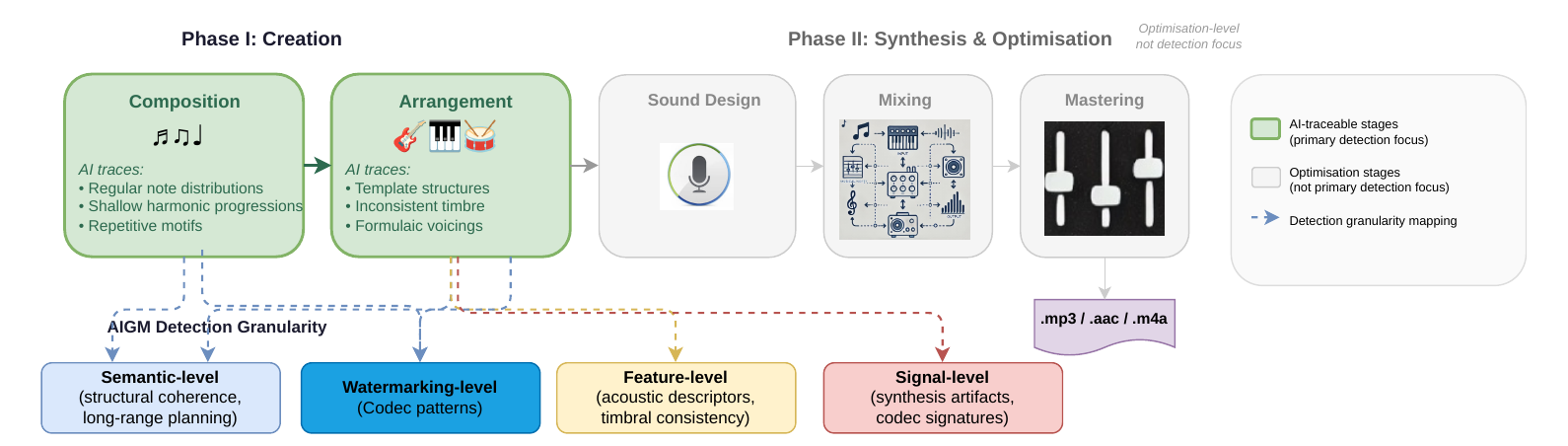}
\caption{Five steps of music production with AIGM detection granularity mapping. \textbf{Composition}, where scattered melodies, harmonies, and rhythms are created to form basic music components; \textbf{Arrangement}, where these components are organised into complete pieces with selection of instruments and harmonic structures. These two stages (in green) generate the core musical content and leave characteristic AI traces such as statistically regular note distributions, structurally shallow harmonic progressions, and formulaic arrangement patterns that are targeted by semantic-level, feature-level, and watermark-level detection methods. \textbf{Sound Design}, where tones are created and modified with synthesisers to achieve sound effects; \textbf{Mixing}, where the balance of each audio track is adjusted for overall coherence; \textbf{Mastering}, the final optimisation such as format preparation for different devices and final check—these post-arrangement stages primarily introduce optimisation-level modifications and fall outside the primary AIGM detection scope. Signal-level detection targets synthesis artifacts that may persist across all stages.}
\label{fig:music production steps}
\end{figure*}

Although AIGM detection, like other audio deepfake detection tasks, is fundamentally a binary classification problem aiming to determine whether a piece of music is human- or AI-generated, several key differences make AIGM detection uniquely challenging. Music, as an art form, is highly subjective, with interpretative elements that can vary widely across audiences and cultures \cite{taylor2022music}. Additionally, music theory introduces unique structures, harmonies, and compositional techniques, offering distinctive features that are not typically present in other audio forms \cite{hernandez2022music}. Furthermore, music often incorporates lyrics with poetic qualities \cite{Irvine2015}, which introduces an additional layer of complexity and shifts AIGM detection tasks into the multimodal domain. These features make AIGM detection a specialised task, requiring tailored ways that account for the complexity and artistry inherent in music.

Therefore, we aim to fill this gap by providing an overview in AIGM detection. There are five essential steps to produce a piece of music: composition, arrangement, sound design, mixing, and master tape production \cite{kwiecien2024technical}, shown in Fig. \ref{fig:music production steps}. We consider the first two phases, composition and arrangement, to have the most creativity and foundational stages in high-quality music production. These steps establish the basis upon which later stages, such as sound design, mixing, and mastering, build. The influence of the first two stages extends further as they shape the core of the music in ways that persist beyond technical refinement. Consequently, the focus in AIGM detection is directed at these initial creative stages. The latter stages, while essential for enhancing the auditory quality, primarily serve as aesthetic enhancements rather than altering the essence of the music. Although we leverage automated tools for them to improve final output quality, they are not the focus of AIGM detection as considered here.

The structure of our paper is as follows: Section \ref{3} focuses on unique music feature understanding methods, specifically those that are fundamental components for AIGM and its detection. Section \ref{2} provides a review of existing AIGM detection tools, as well as foundational and recent audio deepfake detection methodologies, noting areas of overlap with AIGM detection and potential future directions of the latter. Section \ref{4} explores applications of AIGM along with the broader implications for the field. Finally, Section \ref{5} outlines future research directions and key challenges.

In summary, our contributions are as follows: 
\begin{itemize} 
\item To the best of our knowledge, we present the first comprehensive review of AIGM detection methods. 
\item We highlight the importance of focusing on intrinsic music features for AIGM and its detection, moving beyond surface-level techniques like watermark. 
\item We distinguish between audio and multimodal deepfake detection and AIGM detection, discussing potential adaptations and transfers between the domains, a clear pathway from AI audio detection to AIGM detection. 
\item Our work addresses existing gaps in audio deepfake detection reviews by providing comprehensive coverage and critical analysis of each approach. 
\end{itemize}

\section{Paper Taxonomy and Searching methodology }

\begin{figure*}[h]
    \centering
    \includegraphics[width=\linewidth]{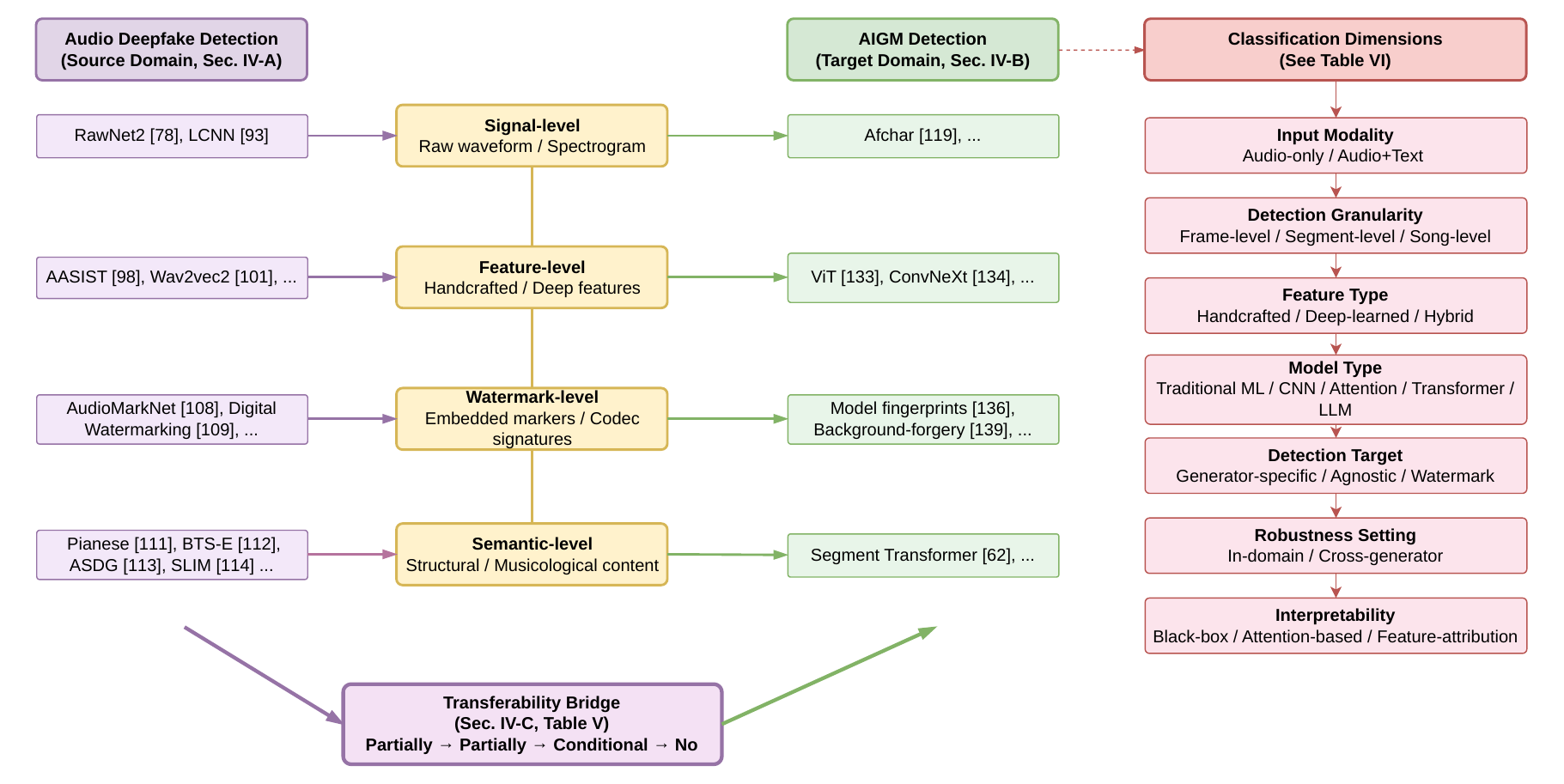}
    \caption{Overview of Musical Features and Taxonomy of AIGM Detection Methods. Key musical features such as melody, harmony, rhythm, and etc. are first summarised, followed by a categorisation of detection approaches, signal, feature, watermark, and semantic levels, highlighting insights transferable from audio deepfake detection.}
    \label{fig:taxomony}
\end{figure*}

As we aim to review the techniques related to AIGM detection, and the number of dedicated studies in this area remains limited, our survey adopts a broader perspective. We first introduce key characteristics of music production and musical structure, which are essential for understanding the distinctive properties of AIGM and identifying potential cues for detection. Subsequently, we review representative approaches in audio deepfake detection, as many of the underlying signal processing and machine learning techniques are transferable to music detection tasks. By combining these perspectives, our taxonomy provides a conceptual foundation for AIGM detection and highlights promising directions for future research in this emerging field. The overview of the taxonomy is shown in more details in Fig. \ref{fig:taxomony}. 

Specifically, we emphasize that the scope of this survey is explicitly bounded to AIGM detection, drawing transferable insights from the two neighbouring domains of audio deepfake detection and multimodal detection. We therefore do not address tasks that, while related, are definitionally distinct from AIGM detection. In particular, we refrain from discussing 
\begin{itemize}
\item 
Plagiarism detection, which asks whether a piece reproduces existing human-composed material rather than whether it is AI-generated,
\item Source attribution, which identifies which specific generative model produced a given output—a qualitatively different task from binary AIGM detection that requires multi-class classification over a potentially open set of generators, and whose methods such as model fingerprinting, embedding-space matching and evaluation protocols such as closed-set vs.\ open-set attribution differ fundamentally from those of binary detection, and
\item Singing voice deepfake detection, targeting AI-generated vocals rather than musical composition as a whole. 
\end{itemize}
Related topics are briefly mentioned in Section \ref{4}.  This boundary of scope follows the current state of the field, in which dedicated AIGM detection has not yet been fully established as a research field, but we nevertheless aim to narrow down the boundaries rather than discussing all neighbouring problems under one umbrella term. In particular, source attribution is excluded because its methodology and evaluation framework differ substantially from binary AIGM detection: attribution requires access to known generator fingerprints or training on generator-specific data, operates over a multi-class label space, and faces the challenge of open-set generalisation to unseen generators—none of which are shared by the binary decision that defines AIGM detection.

The detection task considered throughout Sections~\ref{3} and~\ref{2} is whole-song binary classification: given a complete piece of music, determine whether it is human-composed or fully AI-generated. Real-world music production, however, rarely yields a single global label: partially AI-generated stems, AI-generated accompaniment paired with human vocals, human--AI co-creation, and post-processing operations such as remixing, pitch-shifting, and mastering can render a single binary decision insufficient. We discuss these mixed-authorship and post-processed scenarios as extensions of the binary framework in Section~\ref{4}, and note that the proposed taxonomy is in principle applicable at finer granularities provided datasets and evaluation protocols become available.

To conduct a systematic and comprehensive review of AIGM detection section, we performed keyword-based searches in major scholarly databases, including Google Scholar, IEEE Xplore, and the ACM Digital Library. Search terms combined concepts related to artificial intelligence or AI-generated content (e.g.,``AI-generated,'' ``music deepfake'') with music-related descriptors. Inclusion criteria were studies that focus on AIGM or related audio deepfake detection methods potentially transferable to music tasks. Papers not directly relevant to AIGM, lacking methodological details, or not peer-reviewed were excluded. Selected studies were then grouped according to the categories defined in Section IV to construct a coherent taxonomy.

For audio deepfake detection, we applied a similar search strategy; however, this review is presented in a summarised rather than fully systematic manner. The aim of this section is to provide an overview of representative methods and insights that could inform the development of future music deepfake detection techniques, rather than to offer exhaustive coverage.

\section{Music features and generation}
\label{3}

This section provides an overview of fundamental music components and the core application of AI in generating musical segments, establishing a basis for discussing detection techniques. By emphasising musicological features, we focus on identifying unique characteristics that go beyond basic acoustic frequency features, enabling us to capture the subtleties necessary for distinguishing genuine compositions from AI-generated ones. While prior surveys have reviewed foundational methods across the music field \cite{ma2024foundationmodelsmusicsurvey}, they cover a broad scope without a targeted focus, leaving behind the need for a more specialised analysis. 

Generally speaking, most AI methods for the processing of music, including its generation, are based on an abstraction of music in the form of features. These features are typically categorised into three primary approaches. The first method involves models, which mimic the necessary abstractions that humans apply for composition through methods such as supervised learning, implicitly learning musical patterns representation in embedding space for understanding some music features. For example, Arpita et al.\ design a deep network for genre recognition to understand the features \cite{10421180}. However, it remains challenging to determine whether the model has genuinely internalised meaningful features or simply memorised familiar structures. Another type of approaches conditions the model on specific features, such as generating harmony based on a given melody, which enhances explainability and reliability but requires careful model design. The third, more advanced method aims to integrate musicological understanding, where the model directly comprehends the theoretical aspects of music, offering a more refined and promising direction for AI-driven music generation. Musif, an existing Python package for feature extraction is based on this method \cite{llorens2023musif}. It extracts features based on algorithms designed by musicians, thought it is a bit out-dated. We aim to focus our efforts on advancing models capable of achieving a comprehensive understanding of musicology through the integration of music features. In the following sections, we outline specific levels of music feature abstraction from a musicological perspective, alongside the corresponding AI models designed to interpret these features effectively. The overall summarisation is shown in Table \ref{tab:music_features}.

\paragraph{Content-based features} In the composition stage, four primary features convey content, emotion, and stories: melody, harmony, rhythm, and lyrics.

\textbf{Melody}, a sequence of musical notes organised in a rhythmic pattern \cite{benward2009music}, serves as a prominent feature in musical works. For example, in ``Twinkle, Twinkle, Little Star"\footnote{\url{https://www.loc.gov/item/2023838232/}}, the melody follows the pattern \emph{C C G G A A G}. Melodies are often cohesive and memorable, sometimes forming a recognisable `hook' that can garner widespread appeal. Being a key component in musical theory, the concept of melodies has also been proven a helpful tool in automatic deciphering of music. For instance, deep learning techniques have shown effectiveness in extracting and interpreting melodies from audio, outputting melody feature representation using F0 or feature maps \cite{rao2022melody}. Another particular approach involves leveraging frequency-temporal acoustic features combined with attention networks to enhance melody extraction. This method creates representations enriched with semantic information through weighted feature fusion \cite{yu2021frequency}. Other elements, such as tone and octave, can be incorporated as additional conditions to improve melody comprehension. By rearranging the input representation across frequency bins, researchers generate tone, octave, and salience maps that facilitate downstream fusion processes, achieving strong results on melody recognition metrics such as overall accuracy (OA) and raw pitch accuracy (RPA) \cite{chen2022tonet}.

As a central component of music, melody has shown significant utility in applications such as AIGM. For instance, alignment of melody with raw audio enables its use as a condition in generative diffusion models~\cite{wei2024melody}, which was, for instance, done to create the MusicNet dataset~\cite{wei2024melody}. From a detection perspective, melody represents a promising but underexplored feature for AIGM detection. While melody is a fundamental component of musical composition, it remains unclear which specific melodic properties reliably distinguish AI-generated from human-composed music. Preliminary observations suggest that AI-generated melodies may differ statistically from human compositions in terms of interval distributions and long-range motivic development~\cite{li2019comparison}, but dedicated investigation of melody as a detection signal is lacking and represents an open research question.

\textbf{Harmony} is generally defined as the combination of chords and chord progressions \cite{benward2009music}. It serves as a bridge between melody and song structure, linking musical elements with semantic and structural meanings. For instance, in the song ``Twinkle, Twinkle, Little Star," harmonies such as \emph{C major, G major, or F major} may be used to support the melody. In extended musical forms, structures such as verse-chorus help to organise thematic material, balancing repetition and contrast. The A section (verse) introduces thematic content, while the B section (chorus or refrain) reinforces the main message with a memorable, repeated idea.

Harmony extraction has been an active research area, with early efforts utilising traditional methods such as harmonic change detection functions and distance calculations \cite{Harte2006Detecting}. Recently, advanced approaches have incorporated ontological representations of chord progressions, enhancing the understanding of harmonic relationships \cite{kantarelis2023functional}. Harmony has also proven valuable in music generation, where it is adopted as a conditioning factor in models such as the Harmony-Aware Hierarchical Music Transformer (HAT) \cite{Zhang_2022}.

From a detection standpoint, harmony is particularly diagnostic: current AIGM systems may produce harmonically plausible but structurally shallow progressions, favouring common cadential patterns while often lacking the functional harmonic complexity, such as secondary dominants, modal mixture, or long-range tonal planning, that characterises human composition. This gap between local harmonic plausibility and global harmonic architecture represents a concrete and underexplored signal for AIGM detection.

\textbf{Rhythm} is typically defined as a pattern of sounds and silences that highlights strong and weak beats \cite{benward2009music}. For example, in Twinkle, Twinkle, Little Star, the rhythm often follows a $\frac{4}{4}$ time signature, with the 1st and 3rd beats emphasised as strong beats. Rhythm also influences the speed of musical progression, known as \emph{tempo} \cite{tempo}. Tempo detection serves as a way to represent rhythm and provides a method for deeper rhythm comprehension.

Advanced models for rhythm analysis include cadence detection, where graph neural networks can be used to detect Perfect Authentic Cadence (PAC)~\cite{karystinaios2022cadencedetectionsymbolicclassical}. Additionally, rhythm, alongside melody and harmony, plays a vital role in music generation models such as MrBERT~\cite{li2023mrbert} and MusiConGen~\cite{lan2024musicongen}. From a detection perspective, rhythm reveals characteristic weaknesses of current AIGM systems. While models can reproduce statistically common rhythmic patterns, previous studies have observed that generated music can exhibit overly regular temporal structures and reduced rhythmic variability compared to human performances, particularly over longer contexts \cite{dai2021controllabledeepmelodygeneration, Herremans_2019}. These regularities may form detectable AI fingerprints that feature-level detectors can exploit.

\textbf{Lyrics}, as the name implies, serve as the most direct and natural language medium for expressing content in music. As the majority of the information present in lyrics is accessible in written text form, it can be seen as a different modality to audio, requiring different approaches for processing. Numerous models in Natural Language Understanding (NLU) address this textual modality. However, a holistic understanding of lyrics requires multimodal processing, as elements like emphasis, pauses, and other auditory features, which are comprehensible only when combined with the music, are crucial for fully interpreting the meaning of the lyrics. Notable applications in the music domain include exploring the relationship between lyrics and emotions such as LyEmoBERT \cite{revathy2023lyemobert}. %For instance, , a BERT-based model, enables emotion recognition and lyric recommendation.

In the domain of generation, lyrics creation aligns with established text-to-speech (TTS) methodologies. These include models employing various attention-masking strategies \cite{lei2024songcreator}, large language models (LLMs) \cite{ding2024songcomposer}, and traditional controllable TTS frameworks \cite{zhang2023controllable}. Beyond generation, Lyrics also support detection tasks, including innovation detection with Transformers \cite{balluff2024innovations}, authorship attribution with CNNs \cite{yilmaz2023song}, and AI-generated lyric detection with LLMs \cite{labrak2024detecting}.

Although substantial progress has been made in the generation and analysis of lyrics, challenges remain from both generation and detection perspectives. On the generation side, aligning lyrics with audio, determining the precise timing of singing within the musical progression remains difficult, and AI-generated lyrics often follow overly regular rhyme schemes and generic semantic content that diverges from the stylised, idiosyncratic forms of human songwriting. On the detection side, these regularities are potentially exploitable: LLMs have shown promise in detecting AI-generated lyrics~\cite{labrak2024detecting}, and Frohmann et al.\ demonstrate that combining transcribed lyrics with audio features improves robustness over unimodal approaches~\cite{frohmann2025double}. However, no existing work exploits the \emph{cross-modal} relationship between lyrical content and melodic structure, representing a significant gap for future research.

\paragraph{Decoration-based features}
Beyond content-rich features, musical arrangements involve various processes such as selecting timbre, instrument, main emotion tone, genre, etc. \textbf{Timbre} refers to the unique quality or colour of a sound that distinguishes one sound producer (instrument or human voice) from another, such as a \emph{warm} timbre. \textbf{Instrument} denotes the device used to produce musical sounds, for instance, a \emph{piano}. The main \textbf{emotion tone} is the dominant emotions conveyed by the music, such as \emph{joy or sadness}. \textbf{Genres} categorise music by shared style, such as \emph{jazz} \cite{benward2009music}.

These characteristics not only differentiate between various musical tasks but also serve as classification criteria for the analysis of musical pieces, and as guiding elements in music generation~\cite{huang2024musical, sarmento2023gtr, hung2021emopia}. Critically, from a detection perspective, decoration-based features expose a characteristic limitation of current AIGM systems: while commercial tools can condition generation on genre or instrument labels, the resulting timbral and stylistic consistency is often superficial. AIGM may lack the coherent timbral identity across instruments that pervades an entire human-composed piece, a plausible but as yet empirically unverified hypothesis, making timbre and genre consistency potentially discriminative features for AIGM detection that complement the content-based features.

\paragraph{Traditional music generation}
The content-based and decoration-based features discussed above not only underpin human music composition but also define the design space within which AIGM models operate and where they fall short. \textbf{Foundational generative architectures}, including encoder-decoder models such as the Music Transformer~\cite{dong2023multitrack} and MuseNet, VAE-based models such as MusicVAE~\cite{roberts2018musicvae}, GAN-based models such as MuseGAN~\cite{dong2018musegan}, and diffusion-based models such as DiffWave~\cite{kong2021diffwaveversatilediffusionmodel}, have each demonstrated the capacity to produce locally coherent musical output. \textbf{Commercial platforms} including Jukebox\footnote{\url{https://openai.com/index/jukebox/}}, Suno AI\footnote{\url{https://suno-ai.org/}}, and AIVA\footnote{\url{https://www.aiva.ai/}} extend these capabilities to prompt-conditioned generation at scale, though their underlying architectures and training corpora are not disclosed. \textbf{Conditioned generation approaches} further leverage modality-specific features: text-to-audio methods extract melodic themes from lyrical content~\cite{9327976}, lyrics--melody alignment techniques enforce structural correspondence between the two modalities~\cite{yu2021conditional}, and audio-to-audio methods enable melodic reinterpretation and continuation~\cite{yazawa2014melody, chen2020continuous}, with some models such as SongComposer~\cite{ding2024songcomposer} operating in both directions. Despite this diversity of architectures and conditioning strategies, a consistent limitation emerges across all categories: current AIGM models optimise for local musical plausibility, producing moment-to-moment coherence in melody, harmony, and rhythm, but frequently struggle to replicate the global structural planning that characterises human composition. Properties such as large-scale thematic development, long-range harmonic narrative, and the intentional balance of tension and resolution across an entire piece require compositional intent and musical understanding that current generative architectures do not possess~\cite{taylor2022music, hernandez2022music}. This gap between local plausibility and global coherence is not merely an aesthetic shortcoming: it constitutes a theoretically grounded and architecturally persistent basis for AIGM detection. Crucially, because this deficit reflects a fundamental constraint of current generative models rather than generator-specific artifacts, we hypothesise that detectors exploiting long-range musical structure, as the Segment Transformer~\cite{kim2025segment} begins to do, may generalise more effectively across generators and production pipelines than those relying on surface-level spectral features—though this hypothesis awaits cross-generator empirical validation.

\begin{table*}[t]
\centering
\caption{Summary of musicological features and their potential as AIGM detection signals.}
\label{tab:music_features}
\resizebox{2\columnwidth}{!}{%
\begin{tabular}{@{}llll@{}}
\toprule
\textbf{Category} & \textbf{Feature} & \textbf{Characteristic AI Weakness} & \textbf{Detection Status} \\
\midrule
\multirow{4}{*}{Content}
& Melody & Statistical deviations in interval distributions and long-range motivic development~\cite{li2019comparison} & Underexplored \\
\cmidrule{2-4}
& Harmony & Structurally shallow progressions; lacks functional complexity (secondary dominants, modal mixture, long-range tonal planning) & Underexplored \\
\cmidrule{2-4}
& Rhythm & Overly regular temporal structures; reduced rhythmic variability~\cite{roberts2018musicvae} & Preliminary \\
\cmidrule{2-4}
& Lyrics & Overly regular rhyme schemes; generic semantic content; cross-modal inconsistency with melody~\cite{frohmann2025double} & Preliminary \\
\midrule
\multirow{2}{*}{Decoration}
& Timbre \& Instrument & Incoherent timbral identity across instruments; superficial stylistic consistency & Hypothesised (not empirically verified) \\
\cmidrule{2-4}
& Genre \& Emotion & Superficial genre-level conditioning without deep stylistic coherence & Hypothesised (not empirically verified) \\
\midrule
\multicolumn{2}{l}{Traditional Generation} & Local plausibility vs.\ global structural incoherence; lack of long-range thematic development and harmonic narrative~\cite{taylor2022music, kwiecien2024technical} & Hypothesised (not empirically verified) \\
\bottomrule
\end{tabular}%
}
\end{table*}

\section{Detection Methods}
\label{2}

Having established the musicological features that distinguish AI-generated from human-composed music in Section~\ref{3}, we now review detection methods. Since dedicated AIGM detection research remains nascent, we adopt a structured progression: we first survey the mature domain of audio deepfake detection in Section~\ref{sec:audio_deepfake}, then examine existing AIGM detection methods in Section~\ref{sec:aigm_detection}, and finally analyse the transferability between the two domains to identify a principled pathway from audio deepfake detection to AIGM detection in Section~\ref{sec:transferability}. We conclude this section by discussing multimodal fusion as a complementary strategy in Section~\ref{sec:multimodal}.

%% ============================================================
%% IV.A Audio Deepfake Detection (MATURE SOURCE DOMAIN — FIRST)
%% ============================================================
\subsection{Audio Deepfake Detection}
\label{sec:audio_deepfake}

Audio deepfake detection has developed a rich body of methods over the past decade, producing a wide array of models and benchmark datasets. Although these methods target speech rather than music, many of the underlying signal-processing pipelines and model architectures are potentially transferable to AIGM detection. In this subsection, we review the most foundational and representative methods, as well as a selection of recent advancements that address gaps in prior surveys, organised by detection level. This provides the necessary background for the transferability analysis in Section~\ref{sec:transferability}.

\subsubsection{Datasets}
In the general domain of audio deepfake detection, a vast amount of research has emerged in recent years. The most popular datasets include specialised resources across various subdomains. For pure audio deepfake detection, WaveFake \cite{wavefake2021}, FakeOrReal \cite{fakeorreal}, FakeAVCelebV2 \cite{fakeavcelebv2}, and the ADD series from 2022 and 2023 \cite{add2022, add2023} are prominent datasets. Datasets focusing on spoofing and speaker verification include the ASVspoof series (2015, 2019, and 2021) \cite{asvspoof2015, asvspoof2019, asvspoof2021} and CVoice \cite{li2024safeear}. For speech-focused data, resources like the Conversational AMI-Corpus \cite{ami_corpus}, Baidu Silicon Valley AI Lab dataset \cite{baidu_svail}, and VCC \cite{vcc_dataset} are well established benchmark datasets. Multimodal deepfake detection benefits from datasets such as MLAAD \cite{mlaad}, while environmental sound detection is supported by datasets like sound8k \cite{sound8k} and In-the-Wild \cite{muller2022does}. These datasets cater to a lot of research needs and have been extensively utilised in ongoing studies.

\subsubsection{Signal-Level Methods}
Signal-level approaches operate directly on raw waveforms or minimal time--frequency representations, learning discriminative patterns without explicit feature engineering. Representative methods in this category include end-to-end architectures such as RawNet2~\cite{rawnet22021}, which processes raw waveforms directly, and dual-branch frameworks that combine temporal and spectral analysis~\cite{di2025end, lee2025dual}. While these approaches avoid the bias introduced by handcrafted feature selection, their capacity to generalise across domains remains limited, as the statistical patterns they capture are often generator-specific~\cite{tahaoglu2025deepfake}.

\subsubsection{Feature-Level Methods}
Beyond raw waveform approaches, various representations can facilitate more effective processing at the feature level, which has gained considerable interest in audio classification tasks more generally~\cite{zaman2023survey}. Audio features can be broadly categorised into handcrafted and deep learning-based features.

\paragraph{Handcrafted features} These include perceptual attributes closely aligned with human auditory perception, such as pitch~\cite{xue2022audio}, harmony~\cite{liu2024harmonet}, onset strength, Harmonics-to-Noise Ratio (HNR), intensity, chromagram~\cite{bakken2025deep}, and fundamental frequency (F0), as well as physical frequency-domain representations such as Mel-Frequency Cepstral Coefficients (MFCC)~\cite{jbara2025deepfake}, Mel-spectrograms, logarithmic spectrograms, Linear-Frequency Cepstral Coefficients (LFCC), and the Constant-Q Transform (CQT). Perceptual features capture aspects directly linked to how listeners interpret audio, serving as proxies for vocal naturalness in speech. For instance, pitch contour and HNR reflect the regularity and breathiness of human voice production, which synthetic speech struggles to replicate convincingly~\cite{xue2022audio, liu2024harmonet}. Physical features such as MFCCs and Mel-spectrograms capture timbral envelope information in the frequency domain, and have proven effective at exposing the spectral artifacts introduced by vocoders and neural synthesis pipelines in speech deepfake detection~\cite{jbara2025deepfake}.

\paragraph{Deep learning-based features} These are extracted via pre-trained models~\cite{gohari2025audio, abdulhamied2025deepfake}, and offer task-relevant representations without explicit feature engineering, and have demonstrated strong performance particularly when the pre-training domain matches the detection target, a condition that holds well for speech-to-speech deepfake detection but, as shown in Section~\ref{sec:transferability}, becomes problematic when the target domain shifts to music.

Once features are extracted from audio signals, they are fed into classifiers ranging from traditional statistical models to large-scale neural architectures. Traditional approaches such as Q-SVM~\cite{singh2021detection} and Gaussian Mixture Models (GMM)~\cite{chettri2018deeper} model the statistical distribution of handcrafted feature embeddings and remain useful as interpretable baselines, though their capacity to capture complex temporal dependencies is limited. Hybrid architectures that combine handcrafted front-end features with deep learned embeddings~\cite{wen2022multi} offer improved performance by exploiting the complementary strengths of both feature types. Among deep learning models, convolutional neural networks (CNNs)~\cite{LeCun1998Gradient} have become the dominant paradigm for spectrogram-based detection, operating on the assumption that synthetic speech leaves spatially localised artifacts in the time-frequency domain that convolutional filters can detect. Within this family, architectures vary in the trade-off between efficiency and representational capacity: Light CNN (LCNN)~\cite{lavrentyeva2019stc} prioritises lightweight feature extraction, while RawNet2~\cite{rawnet22021}, ResNet-based~\cite{resnet2015, Li2020}, and SeNet-based~\cite{Senet2019, zhang2023improving} architectures achieve richer representations through greater depth and channel attention mechanisms that suppress irrelevant spectral regions. AASIST~\cite{aasist2021} extends this further by integrating spectro-temporal graph attention networks, explicitly modelling dependencies across both time and frequency axes rather than treating them independently---an architectural choice motivated by the observation that deepfake artifacts often manifest as inconsistencies in the joint time-frequency structure of the signal. More recently, Vision Transformer-based models~\cite{dosovitskiy2021imageworth16x16words, Soudy2024DeepfakeCViT} and Wav2Vec2~\cite{wav2vec22020} leverage self-attention to capture long-range dependencies that local convolutional filters miss, while large language model-based approaches~\cite{gu2025allm4add, li2025dfallm} exploit cross-modal semantic representations for detection. A comparison of these models is shown in Table~\ref{table:deepfake_models}. Comparative studies~\cite{li2022comparative, li2024detectingmachinegeneratedmusicexplainability, yang2024robust, wen2022multi} reveal that no single feature or architecture dominates across all conditions. Performance is highly dependent on the match between training and test distributions, with models relying on generator-specific spectral artifacts degrading substantially under domain shift. It is a limitation that becomes particularly acute in the AIGM setting, where generators are diverse, rapidly evolving, and often undisclosed.

\begin{table*}[htbp]
\caption{Representative Feature-Level Audio Deepfake Detection Methods}
\centering
\begin{tabular}{|p{2.2cm}|p{2.6cm}|p{5.8cm}|p{3.2cm}|p{0.8cm}|}
\hline
\textbf{Category} & \textbf{Method} & \textbf{Key Idea} & \textbf{Typical Features} & \textbf{Year} \\
\hline

Traditional ML & Q-SVM \cite{singh2021detection} & Uses quadratic support vector machines to classify spectral patterns in audio signals, demonstrating effectiveness on \textbf{high-dimensional handcrafted features}. & Spectrogram, Mel-spectrogram & 2021 \\
\hline

Traditional ML & GMM-based detection \cite{chettri2018deeper} & Models the \textbf{statistical distribution of acoustic features} to distinguish real and synthetic speech, often used as a classical baseline in spoofing detection tasks. & CQCC, LFCC & 2018 \\
\hline

CNN-based & LCNN \cite{lavrentyeva2019stc} & \textbf{Lightweight convolutional} neural network designed for spoofing detection, capable of extracting discriminative spectral patterns from audio representations. & Spectrogram, LFCC & 2019 \\
\hline

CNN-based & RawNet2 \cite{rawnet22021} & End-to-end deep neural network that directly processes raw waveforms to learn discriminative representations for detecting synthetic speech. This approach is \textbf{commonly adopted because audio signals are often represented as spectrograms}, which can be interpreted as time–frequency images. & Raw waveform & 2021 \\
\hline

CNN-based & ResNet-based detectors \cite{resnet2015,Li2020} & Utilises residual convolutional architectures to learn \textbf{hierarchical representations} of audio spectrograms for robust deepfake detection. & Spectrogram, Log-Mel & 2022 \\
\hline

Attention-based & SE-ResNet / SeNet \cite{Senet2019,zhang2023improving} & Introduces \textbf{channel attention mechanisms} to emphasise informative spectral features and suppress irrelevant noise. & STFT, Spectrogram & 2023 \\
\hline

Attention-based & AASIST \cite{aasist2021} & Integrates spectro-temporal attention mechanisms with \textbf{graph attention networks} to model artifacts in spoofed speech signals. & Raw waveform & 2021 \\
\hline

Transformer-based & Vision Transformer-based detection \cite{dosovitskiy2021imageworth16x16words, Soudy2024DeepfakeCViT} & Treats spectrograms as image patches and applies transformer architectures to \textbf{capture long-range dependencies} in audio representations. & Mel-spectrogram, Spectrogram & 2024 \\
\hline

Transformer-based & Wav2Vec2-based detection \cite{wav2vec22020} & Utilises \textbf{pretrained self-supervised} speech representations to capture high-level acoustic characteristics for deepfake detection. & Raw waveform & 2020 \\
\hline

LLMs-based models & Audio-LLM based detection \cite{gu2025allm4add} & Explores \textbf{large-scale pretrained audio-language models} for deepfake detection through cross-modal representation learning. & Raw waveform & 2025 \\

\hline
\end{tabular}
\label{table:deepfake_models}
\end{table*}

\subsubsection{Watermark Methods}
Building on the feature-level methods above, recent research has introduced approaches that target traces left by generative models. We distinguish passive artifact detection, which identifies inadvertent statistical traces left by generation processes without deliberate embedding, from active watermark, which exploits markers deliberately embedded by cooperative generators. In audio deepfake detection, passive artifact detection is the more common approach, as adversarial generators cannot be assumed to cooperate; active watermark applies primarily in compliance and copyright-protection scenarios where the generation pipeline embeds detectable signals by design.

At the watermark level, detection relies on post-hoc analysis of the audio signal itself, identifying imperceptible statistical traces, such as vocoder artifacts or characteristic frequency patterns, inadvertently left by generative models during synthesis~\cite{cao2025watermarking}. Techniques such as AudioMarkNet~\cite{zong2025audiomarknet} and digital watermark detection~\cite{qureshi2021detecting} exploit these systematic traces to confirm synthetic origin. However, this paradigm is inherently fragile: as generative models improve and leave fewer detectable artifacts, passive detection degrades accordingly~\cite{malik2019fighting}, making it an insufficient standalone solution for open-world detection where the generator is unknown and evolving.

\subsubsection{Semantic-Level Methods}
Beyond watermark, semantic-level methods shift the focus to higher-order relationships and contextual dependencies within the audio. These methods aim to capture the nuances of linguistic, stylistic, and structural patterns that are difficult for generative models to replicate. For instance, Pianese et al. \cite{Deepfakeaudiodetection} proposed a similarity-based clustering algorithm for speaker verification, assessing audio authenticity through semantic patterns rather than raw signal features. Similarly, the Breathing-Talking-Silence Encoder (BTS-E) \cite{doan2023bts} leverages breathing and silence as distinguishing features, taking advantage of the fact that AI-generated audio struggles to replicate these low-information segments. The Aggregation and Separation Domain Generalisation (ASDG) model \cite{xie2023domain} implicitly focuses on feature variations by grouping real audio and categorising fake songs across different domains. Style-focused models like the Style-Linguistics Mismatch (SLIM) \cite{zhu2024slim} capture dependencies between linguistic and stylistic attributes, helping identify inconsistencies that suggest AI generation. The SafeEar model \cite{li2024safeear}, which obfuscates content to protect privacy during detection, also belongs to this category, as it focuses on understanding the deeper semantic structure of audio content and preserving its integrity during detection. Additionally, Jellali et al. \cite{jellali2025pushing} proposed a hybrid approach that combines MFCC and spectral contrast, capturing more advanced audio features for more robust detection. Chen et al. \cite{chen2025region} introduced region-based optimization in continual learning for audio deepfake detection, focusing on adapting models to new, unseen data. Moreover, Dilbar et al. \cite{dilbar2025audiofakenet} developed Audiofakenet, a model for reliable speaker verification in audio deepfake, using advanced verification techniques to ensure authenticity and improve detection accuracy.

We critically reviewed these works due to their highly representative methods compared to other non-peer-reviewed research. While there are other recent works, we have selectively emphasised these and summarised them in Table \ref{tab: Recent Audio Deepfake Detection Models}. Opportunities remain to innovate new models through the fusion of acoustic, physical, and perceptual features. Furthermore, enhancing model explainability and improving the robustness of generalisation are possible future work directions.

\begin{table*}[htbp]
\centering
\caption{Summary of Watermark- and Semantic-Level Audio Deepfake Detection Models}
\label{tab: Recent Audio Deepfake Detection Models}
\begin{tabular}{|p{0.8cm}|p{2.5cm}|p{3.8cm}|p{3.8cm}|p{1.8cm}|p{0.5cm}|}
\hline
\textbf{Model} & \textbf{Architecture} & \textbf{Advantages} & \textbf{Limitations} & \textbf{Baseline Comparison} & \textbf{Year}\\ \hline
Qureshi et al.\cite{qureshi2021detecting} & Digital Watermark & Focuses on watermark for detecting AI-generated content & Not scalable to all types of audio & GMM, CNN & 2021 \\ \hline
Pianese et al.\cite{Deepfakeaudiodetection} & Centroid-Based, Multi-Similarity & Proposes novel speaker ID identification methods; & Speaker recognition is inherently difficult & ResNet, Cosine Similarity & 2022\\ \hline
Doan et al.\cite{doan2023bts} & CNN2S2, RNN & Incorporates breathing, talking, and silence detection, challenging generators & Limited comparisons; human speech variation in breath and silence needs analysis & ResNet2 & 2023 \\ \hline
Xie et al.\cite{xie2023domain} & LCNN & Strong baseline comparisons; innovative real-fake separation in music & Variability in real audio samples may impact results & GMM, Inception-based, ResNet, RawNet, RawGAT-ST, AASIST & 2023 \\ \hline
Li et al.\cite{li2024safeear} & Residual Vector Quantisers (RVQs), Encoder-Decoder, Threat Model & Introduces a dataset; uses decoupling for privacy; demonstrates security & Privacy retention unclear; lack of mathematical validation & AASIST, Rawformer, Wav2Vec2 & 2024 \\ \hline
Zhu et al.\cite{zhu2024slim} & MLP & New method with unique features; high interpretability & Misclassifies speech with unusual style-linguistic dependencies & AASIST & 2024 \\ \hline
Zong et al.\cite{zong2025audiomarknet} & CNN & Detects audio deepfake using embedded watermarks & Dependence on watermark embedding quality & RawNet, Wav2Vec2 & 2025 \\ \hline
Jellali et al.\cite{jellali2025pushing} & Hybrid MFCC, Spectral Contrast & Pushes boundaries by combining MFCC and spectral contrast for robust detection & May require complex feature extraction & RawNet, Wav2Vec2 & 2025 \\ \hline
Chen et al.\cite{chen2025region} & Region-Based Optimization, Continual Learning & Region-based optimization improves continual learning for deepfake detection & Limited analysis of real-world applications & CNN, ResNet, Wav2Vec2 & 2025 \\ \hline
Dilbar et al.\cite{dilbar2025audiofakenet} & Speaker Verification & Reliable speaker verification for deepfake detection & Needs further model optimization for efficiency & GMM, ResNet & 2025 \\ \hline
\end{tabular}
\end{table*}

\subsubsection{Limitations}
Though the progress in this field is notable, audio deepfake detection faces two interconnected limitations. First, while handcrafted features such as MFCCs, LFCC, and Mel-spectrograms are stable extraction methods, their selection remains largely empirical: comparative studies show that no single feature consistently outperforms others across datasets and generators~\cite{li2022comparative, yang2024robust}, and there is no principled framework for determining which acoustic properties carry genuine discriminative power for detecting AI-generated audio. Second, regardless of which features are selected, detectors learn to identify generator-specific artifacts such as watermark that change or disappear entirely when a new generative architecture is encountered. Newer flow-matching and neural codec-based systems generate audio through fundamentally different mechanisms, rendering trained detectors ineffective~\cite{muller2024harder}. In open-world benchmarks, this manifests as EER increases of 10--40 percentage points on unseen generators\footnote{\url{https://www.emergentmind.com/topics/deepfake-eval-2024}}. These limitations are shared with, and indeed amplified in, AIGM detection, motivating the examination of existing AIGM methods and the transferability analysis.

%% ============================================================
%% IV.B AIGM Detection (EMERGING TARGET DOMAIN — SECOND)
%% ============================================================
\subsection{AIGM Detection}
\label{sec:aigm_detection}

Both Afchar \cite{afchar2024detecting} and Cooke \cite{cooke2024good} observe that, although detection models can achieve high accuracy within specific domains, their performance degrades considerably in out-of-domain scenarios. This decline in accuracy highlights the inherent complexity of the task and underscores the need for more robust models capable of capturing deeper, more intricate features. Consequently, few dedicated AIGM detectors have been developed. We review the available datasets and existing detection methods, organised by the same detection-level taxonomy introduced above.

\subsubsection{Datasets}
Currently, only two datasets are specifically designed for AIGM detection: FakeMusicCaps \cite{comanducci2024fakemusiccaps} and SONICS \cite{rahman2024sonics}, with the latter yet to be publicly released. FakeMusicCaps is structured to distinguish the authorship of music, whether created by humans or generated by AI. Its human-made content is derived from the MusicCaps dataset \cite{agostinelli2023musiclmgeneratingmusictext}, while AI-generated content is prompted from MusicCaps captions using five different large language models (LLM). SONICS, in contrast, is a multimodal dataset that contains lyrics and melodies. Human-produced songs are sourced from the Genius Lyrics Dataset\footnote{\url{https://www.kaggle.com/datasets/carlosgdcj/genius-song-lyrics-with-language-information}}, with accompanying audio retrieved from YouTube. AIGM in SONICS utilises technologies such as Suno AI\footnote{https://suno.com} for melody generation and GPT models for lyrics. A detailed comparison of these datasets is presented in Table \ref{table:music_datasets}.

Although dedicated AIGM detection datasets remain scarce, several public datasets in music analysis and AIGM exist, albeit without specific gathering and labelling for AIGM detection. Notable recent examples of datasets containing human-generated music include DALI \cite{Meseguer-Brocal2018DALI}, which provides aligned melodies and lyrics, FMA \cite{defferrard2017fmadatasetmusicanalysis}, a genre-diverse dataset, MAESTRO \cite{hawthorne2018enabling} focused on piano performances, MSD \cite{millionsong}, which offers metadata on one million songs, MusicNet \cite{thickstun2017learning}, intended for instrument recognition, and MTG-Jamendo \cite{bogdanov2019mtg}, which includes classification of instruments, mood, and genre.

In the realm of AIGM, datasets such as SunoCaps \cite{civit2024sunocaps} offer emotional annotations, while the AI Song Contest\footnote{https://www.aisongcontest.com/the-2024-finalists} provides around ten finalised AI-assisted songs per year as part of a music creation competition. Furthermore, Afchar's deepfake dataset \cite{afchar2024detecting} offers a number of songs for study based on FMA, though not published yet. However, a newly introduced dataset, M6 \cite{li2026m6}, has been made available. It covers diverse AI model sources and human-generated music to address data scarcity. Table~\ref{table:music_datasets} summarises these datasets.

\begin{table}[htbp]
\caption{Music Datasets with AI and Human-Generated Content}
\centering
\resizebox{\columnwidth}{!}{
\begin{tabular}{|p{1.8cm}|c|r|c|c|c|}
\hline
\textbf{Dataset} & \textbf{Source} & \textbf{Clips} & \textbf{Modality} & \textbf{Year} & \textbf{Genre} \\
\hline
FMA & Human & 106,574 & Audio & 2017 & Diverse \\
DALI & Human & 5,358 & Audio\&Text & 2019 & Karaoke \\
MAESTRO & Human & 1,276 & Audio & 2018 & Piano \\
MSD & Human & 1 million & Audio & 2011 & Diverse \\
MusicNet & Human & 330 & Audio & 2016 & Classical \\
MTG-Jamendo & Human & 55,000 & Audio & 2019 & Diverse \\
SunoCaps & AI & 256 & Audio & 2024 & Diverse \\
Afchar's & AI & 250,000 & Audio & 2024 & Diverse \\
AISongContest & AI & 10 & Audio & 2024 & Diverse \\
FakeMusicCaps & Mixed & 33,105 & Audio & 2024 & Diverse \\
SONICS & Mixed & 97,164 & Audio\&Text & 2024 & Diverse \\
M6 & Mixed & 13,493 & Audio\&Text & 2026 & Diverse \\
\hline
\end{tabular}}
\label{table:music_datasets}
\end{table}

\subsubsection{Signal-Level Detection}
Signal-level detection learns discriminative patterns directly from raw audio signals or time--frequency representations such as spectrograms. For example, the Spectro-Temporal Tokens Transformer (SpecTTTra)~\cite{rahman2024sonics} processes two-dimensional Mel-spectrograms using a 1D convolutional tokenizer combined with a Transformer encoder, capturing long-range temporal dependencies. While this approach achieves competitive in-domain performance, it operates at a low representational level and is therefore vulnerable to distribution shift: changes in dataset or post-processing models can alter the spectral statistics the model relies upon without changing the underlying musical content, causing performance to degrade on out-of-domain examples~\cite{afchar2024detecting}.

\subsubsection{Feature-Level Detection}
Feature-level detection relies on handcrafted or deep-learned acoustic descriptors, including spectral, prosodic, or pre-trained model-derived features. Comparative analyses of deep learning models on curated AIGM datasets~\cite{wie2025comparative} demonstrate that feature-level representations improve classification over raw signal approaches. Vision Transformer~\cite{dosovitskiy2020image} and ConvNeXt~\cite{liu2022convnext} achieve strong performance by treating Mel-spectrograms as two-dimensional images and applying patch-based or convolutional processing to capture both local texture and global layout, which manifest as systematic patterns in the time-frequency domain. However, this image-based framing also reveals a fundamental assumption: that the discriminative signal lies in \emph{what the spectrogram looks like} rather than \emph{what the music does}. As a consequence, these models are effective when trained and tested on the same generator but struggle to generalise as it is hard to select the optimal feature set for detection. Studies employing EEG-informed emotion embeddings~\cite{martin2025eeg} represent an attempt to incorporate higher-level perceptual features, but stop short of the music-theoretic representations, harmonic structure, rhythmic organisation, long-range form, that would be needed for generator-agnostic detection.

\subsubsection{Watermark Detection}
Watermark detection identifies artifacts or watermarks introduced by specific generative models during synthesis, which encompasses both passive and active approaches. In the AIGM detection context, most methods fall under passive artifact detection, identifying inadvertent fingerprints—systematic biases embedded by generation algorithms rather than deliberately embedded markers. Active watermark, where generators cooperatively embed detectable signals, is often seens in copy-right related application such as AudioSeal \cite{roman2024proactivedetectionvoicecloning} or SynthID\footnote{\url{https://deepmind.google/models/synthid/}}.

For this purpose, recent work proposes detection mechanisms that target the intrinsic fingerprints left by generative models across text, visual, and audio modalities~\cite{cao2025watermarking, cao2025practical}, capturing systematic biases embedded by generation algorithms rather than surface-level acoustic patterns~\cite{afchar2025ai}. Background-forgery detection techniques similarly target subtle signal-level manipulations introduced during AIGM synthesis~\cite{wei2025voices}. While these approaches offer a principled basis for detection, their applicability is inherently conditional: they presuppose that the generative model leaves consistent, detectable traces---an assumption that breaks down when generators are updated, when outputs are post-processed, or when watermarks are deliberately removed~\cite{malik2019fighting}.

\subsubsection{Semantic Consistency Detection}
Semantic consistency detection examines higher-level musical structure to determine whether the content exhibits inconsistencies indicative of AI generation. The Segment Transformer~\cite{kim2025segment} examines inter-segment musical structure to detect inconsistencies indicative of AI generation, representing the most musically grounded approach to date. By modelling relationships across musical segments rather than local spectral windows, it targets precisely the global structural deficit identified in Section~\ref{3}: the inability of current generative models to sustain large-scale thematic coherence and harmonic narrative. This makes semantic-level detection a theoretically motivated approach that we hypothesise may be more robust to generator variation, provided that the structural limitations it exploits indeed reflect architectural constraints shared across generative models rather than model-specific artifacts, a claim that remains to be empirically established. However, this is challenging to train as learning semantic information requires huge amounts of human annotation to achieve consistency of human values and tastes of beauty.

\subsubsection{Limitations}
Despite these advances, current AIGM detectors share three interconnected limitations. First, feature selection remains largely empirical, as handcrafted acoustic features, such as MFCCs, lack the basis for determining which properties carry genuine discriminative power for AIGM detection. Comparative studies confirm this limitation by showing that no single feature consistently outperforms others across generators and datasets~\cite{li2022comparative, yang2024robust}. Second, the music-theoretic features identified in Section~\ref{3} remain largely unexploited by existing detectors, representing a significant gap between what musicology offers as discriminative signals and what current methods actually use. Together, these cause poor generalisation to unseen domains. Because detectors did not learn musically meaningful properties shared across generators, detection accuracy drops substantially when evaluated on out-of-distribution generators~\cite{afchar2024detecting, cooke2024good}.

%% ============================================================
%% IV.C Transferability Analysis (PATHWAY — THIRD)
%% ============================================================

\begin{table*}[ht]
\centering
\caption{Transferability Analysis of Audio Deepfake Detection Methods to AIGM Detection}
\label{tab:transferability}
\resizebox{\textwidth}{!}{%
\begin{tabular}{|l|p{2.5cm}|c|p{2.2cm}|p{2.2cm}|p{2.2cm}|p{2.2cm}|p{2.2cm}|p{2.2cm}|p{2.2cm}|p{2.2cm}|}
\hline
\textbf{Detection Level} & \textbf{Key Assumption} & \textbf{Transferability Holds?} & \textbf{Target Signal Type} & \textbf{Generation Mechanism (audio vs.\ music)} & \textbf{Reusable Architecture} & \textbf{Transferable Cues} & \textbf{Non-transferable Cues} & \textbf{Music-specific Cues} & \textbf{Valid Transfer Conditions} & \textbf{Expected Failure Modes} \\
\hline
Signal-level & Forgery artifacts are visible in raw waveforms or spectrograms & Partially & Raw waveform and spectrogram & Neural vocoder vs.\ diffusion/codec: different synthesis architectures leave distinct waveform traces & RawNet2~\cite{rawnet22021}, LCNN~\cite{lavrentyeva2019stc} & Spectral discontinuities; phase inconsistencies; statistical anomalies & Formant structure; voicing patterns & Dense harmonic overtones; polyphonic overlap; wide dynamic range & In-domain or music-finetuned corpora & Generator-specific artifacts fail to generalise across unseen synthesizers; polyphonic overlap masks low-level synthesis traces \\
\hline
Feature-level & Acoustic and prosodic features are discriminative for fake detection & Partially & Acoustic feature vectors & TTS/VC vs.\ symbolic$+$neural generators: generation distorts feature distributions differently across domains & AASIST~\cite{aasist2021}, ResNet-based~\cite{Li2020} & Temporal modeling principles; discriminative training paradigms & Speaking rate; pitch contour shape; formant trajectories; phonotactics & Melodic contour; harmonic progressions; rhythmic patterns; timbral signatures & Domain-adapted features on music data; cross-domain pre-training & Speech-pretrained encoders encode linguistic priors misaligned with music; prosodic features produce false positives on musical variation \\
\hline
Watermark-level & Generative models embed detectable markers in output & Conditional & Encoded representations & Watermarked TTS vs.\ Watermarked text-to-music or commercial methods with unknown embedding policies & AudioMarkNet~\cite{zong2025audiomarknet} & Watermark detection infrastructure; extraction pipelines & Cooperative-generator assumption & Music generation model enCodec watermark and codec-steganographic signatures & Watermark present unchanged and intact & Watermark removal or degradation; non-compliant generators \\
\hline
Semantic-level & Linguistic naturalness and speaker consistency serve as detection cues & No & Structural and compositional content & LLM-TTS vs.\ text-to-music: both locally plausible but coherence criteria differ—linguistic vs.\ musicological & Segment Transformer~\cite{kim2025segment} & None directly & Linguistic coherence; speaker identity consistency; breathing/silence patterns & Tonal harmony; cadential resolution; long-range form; genre conventions & Music-theoretic ground truth required & No musicological prior; speech-semantic misapplication \\
\hline
\end{tabular}%
}
\end{table*}

\subsection{From Audio Deepfake Detection to AIGM: A Transferability Analysis}
\label{sec:transferability}

As established in Section~\ref{sec:audio_deepfake}, dedicated AIGM detectors remain scarce, making it necessary to draw upon the more mature field of audio deepfake detection. However, music differs fundamentally from speech in its structural complexity, harmonic organisation, and compositional intent~\cite{hernandez2022music, taylor2022music}. Blindly transferring audio deepfake detection methods to AIGM risks inheriting assumptions that do not hold in musical contexts. To address this, we propose a \emph{pathway} that serves three complementary roles: (1)~a \textbf{methodological roadmap} guiding the step-wise adaptation of existing detection techniques, (2)~a \textbf{taxonomy of reusable components} identifying which architectures, features, and training paradigms transfer across domains, and (3)~a set of \textbf{research hypotheses} specifying the conditions under which transfer is meaningful and the failure modes expected when assumptions break down. We analyse each detection level introduced in Fig.~\ref{fig:taxomony}, summarising the results in Table~\ref{tab:transferability}. We also classify the models with different dimensions defined in Fig.~\ref{fig:taxomony} in Table \ref{tab:classification}.

\begin{table*}[ht]
\centering
\caption{Classification of Important Detection Methods by Multi-Dimensional Perspectives. Gen. represents Generator, WM represents watermark, attr. represents attributes}
\label{tab:classification}
\resizebox{\textwidth}{!}{%
\begin{tabular}{|l|c|c|c|c|c|c|c|c|}
\hline
\textbf{Method} & \textbf{Level} & \textbf{Input Modality} & \textbf{Granularity} & \textbf{Feature Type} & \textbf{Model Type} & \textbf{Detection Target} & \textbf{Robustness} & \textbf{Interpretability} \\
\hline
\multicolumn{9}{|l|}{\textit{Signal-level Detection}} \\
\hline
RawNet2~\cite{rawnet22021} & Signal & Audio & Frame & Deep-learned & CNN & Gen.-agnostic & Cross-gen. & Black-box \\
\hline
LCNN~\cite{lavrentyeva2019stc} & Signal & Audio & Frame & Deep-learned & CNN & Gen.-specific & In-domain & Black-box \\
\hline
Afchar~\cite{afchar2024detecting} & Signal & Audio & Frame & Deep-learned & Various & Gen.-specific & In-domain & Black-box \\
\hline
\multicolumn{9}{|l|}{\textit{Feature-level Detection}} \\
\hline
Q-SVM~\cite{singh2021detection} & Feature & Audio & Frame & Handcrafted & Trad.\ ML & Gen.-specific & In-domain & White-box \\
\hline
GMM~\cite{chettri2018deeper} & Feature & Audio & Frame & Handcrafted & Trad.\ ML & Gen.-specific & In-domain & White-box \\
\hline
AASIST~\cite{aasist2021} & Feature & Audio & Frame & Deep-learned & Attention+GNN & Gen.-agnostic & Cross-gen. & Attention-based \\
\hline
ResNet-based~\cite{resnet2015} & Feature & Audio & Frame & Deep-learned & CNN & Gen.-specific & In-domain & Black-box \\
\hline
SE-ResNet~\cite{Senet2019} & Feature & Audio & Frame & Deep-learned & CNN+Attn. & Gen.-specific & In-domain & Attention-based \\
\hline
Wav2Vec2~\cite{wav2vec22020} & Feature & Audio & Segment & Deep-learned & Transformer & Gen.-specific & In-domain & Black-box \\
\hline
ViT~\cite{dosovitskiy2021imageworth16x16words} & Feature & Audio & Segment & Deep-learned & Transformer & Gen.-specific & In-domain & Attention-based \\
\hline
ConvNeXt~\cite{liu2022convnext} & Feature & Audio & Segment & Deep-learned & CNN & Gen.-specific & In-domain & Black-box \\
\hline
EEG-informed~\cite{martin2025eeg} & Feature & Audio & Segment & Deep-learned & Hybrid & Gen.-specific & In-domain & Black-box \\
\hline
\multicolumn{9}{|l|}{\textit{Watermark-level Detection}} \\
\hline
AudioMarkNet~\cite{zong2025audiomarknet} & Watermark & Audio & Frame & Deep-learned & CNN & WM-dep. & In-domain & Black-box \\
\hline
Watermark~\cite{cao2025watermarking} & Watermark & Audio & Frame & Deep-learned & Various & WM-dep. & In-domain & Black-box \\
\hline
Background-forgery~\cite{wei2025voices} & Watermark & Audio & Frame & Deep-learned & CNN & Gen.-specific & In-domain & Black-box \\
\hline
SafeEar~\cite{li2024safeear} & Watermark & Audio & Frame & Deep-learned & Enc.-Dec. & WM-dep. & In-domain & Black-box \\
\hline
\multicolumn{9}{|l|}{\textit{Semantic-level Detection}} \\
\hline
BTS-E~\cite{doan2023bts} & Semantic & Audio & Segment & Hybrid & CNN+RNN & Gen.-specific & In-domain & White-box \\
\hline
SLIM~\cite{zhu2024slim} & Semantic & Audio & Segment & Deep-learned & MLP & Gen.-specific & In-domain & Feature-attr. \\
\hline
Pianese~\cite{Deepfakeaudiodetection} & Semantic & Audio & Segment & Deep-learned & Similarity & Gen.-agnostic & Cross-gen. & Feature-attr. \\
\hline
ASDG~\cite{xie2023domain} & Semantic & Audio & Segment & Deep-learned & LCNN & Gen.-agnostic & Cross-gen. & Black-box \\
\hline
Seg.\ Transformer~\cite{kim2025segment} & Semantic & Audio+Text & Song & Deep-learned & Transformer & Gen.-agnostic & Cross-gen. & Attention-based \\
\hline
Audio-LLM~\cite{gu2025allm4add} & Semantic & Audio+Text & Segment & Deep-learned & LLM & Gen.-agnostic & Cross-gen. & Black-box \\
\hline
\end{tabular}%
}
\end{table*}

\subsubsection{Signal-Level Transferability}
At the signal level, the core assumption is that artifacts or statistical distribution differences in raw waveforms or spectrograms could be detected. This assumption \emph{partially holds} for AIGM: neural vocoders in TTS systems and diffusion/codec-based music generators both leave low-level synthesis traces, and recent work on audio transfer learning via Spectrogram Transformers~\cite{cappellazzo2024parameterefficienttransferlearningaudio} demonstrates the feasibility of adapting signal-level methods across audio domains. However, the generation mechanisms differ. Neural vocoders produce speech-specific formant artifacts, whereas music generators face dense harmonic overtones and polyphonic overlap. This means music spectra are substantially richer than speech, exhibiting a wide dynamic range. Signal-level detectors trained on speech corpora such as ASVspoof~\cite{asvspoof2019} may therefore fail to generalise, as the artifact patterns they rely upon differ in musical contexts~\cite{afchar2024detecting}. Transfer is therefore valid only under in-domain or music-finetuned conditions; on unseen generators, generator-specific artifacts fail to generalise, and polyphonic overlap masks low-level synthesis traces.

\subsubsection{Feature-Level Transferability}
At the feature level, audio deepfake detectors commonly exploit two categories of features: handcrafted acoustic features and deep learned representations as previously stated. In the context of AIGM, both categories require adaptation. Although these features such as pitch are shared between speech and music, their statistical distributions differ across domains due to distinct generative and functional constraints: TTS (text to speech)/VC (voice conversion) models focus on speech prosody (speaking rate, formant trajectories, phonotactics), while symbolic with neural music generators pay attention to melodic contour, harmonic progressions, and rhythmic patterns, leading to different feature behaviors in downstream detection tasks~\cite{ozaki2024globally}. These should therefore be augmented with content-based music features as well as decoration-based features such as timbral signatures, instrument identity, and genre characteristics~\cite{benward2009music}. For deep learned features, models pre-trained on speech corpora encode linguistic priors misaligned with musical structure~\cite{afchar2024detecting}, causing prosodic features to produce false positives on musical variation. Re-training or fine-tuning on music corpora, or adopting music foundation models, would therefore be necessary to capture both the content-level and decoration-level properties of music. Despite these adaptation requirements, the underlying model architectures, such as CNNs, Transformers, and graph attention networks, remain applicable once the feature space is appropriately redefined, making feature-level transfer the most architecturally promising direction.

\subsubsection{Watermark-Level Transferability}
At the watermark level, transferability is \emph{conditional} rather than architectural. Watermark-based detection methods such as AudioMarkNet~\cite{zong2025audiomarknet} are effective when the generative model embeds detectable markers in its output. However, while some TTS systems embed watermarks, many commercial AIGM platforms such as Suno AI and AIVA do not disclose whether such markers are embedded, and watermarks can be deliberately removed or degraded~\cite{malik2019fighting}. Music-specific cues such as text-to-music model watermark and codec-steganographic signatures offer a partial alternative, but these depend on the specific codec architecture and are not universal. Transfer is valid only when watermarks are present unchanged and intact from audio generation model to music; otherwise, watermark removal, degradation, or non-compliant generators render these approaches ineffective. Watermark-based approaches therefore offer a useful but fragile layer of detection, unsuitable as a standalone solution for AIGM.

\subsubsection{Semantic-Level Transferability}
At the semantic level, the transferability gap is most pronounced. Semantic-level audio deepfake detectors exploit cues such as linguistic coherence, speaker identity consistency, and the naturalness of breathing and silence patterns~\cite{doan2023bts, zhu2024slim}. None of these cues translate directly to music. Both LLM-based TTS and text-to-music models produce locally plausible outputs, but their coherence criteria differ fundamentally: speech demands linguistic and speaker consistency, whereas music demands tonal harmony, cadential resolution, long-range formal structure, and genre-specific conventions. This level therefore \emph{does not transfer} and demands entirely new detection frameworks grounded in musicology, of which the Segment Transformer~\cite{kim2025segment} represents an early example. We argue this is where the most impactful and original contributions to AIGM detection remain to be made.

Taken together, transferring from audio deepfake detection to AIGM detection is neither straightforward nor uniform, depending critically on the detection level, generation mechanisms, and domain-appropriate training data. Signal-level methods offer the most immediate transfer potential under constrained conditions; feature-level methods require deliberate re-engineering of the input representation but retain architectural compatibility; watermark-based approaches carry inherent fragility; and semantic-level detection demands domain-specific innovation. This stratified view directly informs the benchmark evaluation criteria discussed in Section~\ref{4}, and the open questions it raises, particularly at the semantic level, motivate the future research directions outlined in Section~\ref{5}. Rather than treating the two domains as overlapping by observation, this analysis provides a concrete, testable framework for determining when and how methods from audio deepfake detection can be meaningfully adapted for AIGM detection.

%% ============================================================
%% IV.D Multimodal Fusion (UNCHANGED)
%% ============================================================
\subsection{Multimodal Fusion (a Complementary Detection Strategy)}
\label{sec:multimodal}

The transferability analysis reveals that single-modality approaches face inherent limitations when applied to AIGM detection, particularly at the semantic level. Music, however, is inherently multimodal, as it integrates melody as an audio modality and lyrics as a textual modality, each carrying complementary information about compositional intent and structure. This multimodal nature of music is not merely a challenge but also an opportunity. Inconsistencies between AI-generated audio and lyrics, or between melodic structure and lyrical content, may serve as discriminative cues that single-modality detectors cannot exploit. Dedicated frameworks for audio-text-based AIGM detection, however, remain absent, and existing research on multimodal deepfake detection is largely confined to speech and video domains. We therefore review foundational multimodal fusion strategies and their application in music research, identifying concrete directions for adaptation to AIGM detection.

The choice of fusion strategy for AIGM detection is not arbitrary but is constrained by the structural properties of music and the inherent limitations of unimodal approaches. Audio-only detectors, despite achieving high in-domain accuracy, fail to generalise to unseen generators and are vulnerable to perturbations such as added noise or pitch shifting~\cite{afchar2024detecting, frohmann2025double}. Lyrics-only detectors, by contrast, require clean and accurately formatted text that is unavailable in practice when only audio is accessible~\cite{frohmann2025double}. These complementary failure modes motivate a multimodal approach in which melody and lyrics are jointly analysed. In the broader deepfake detection literature, cross-modal inconsistency has been established as a reliable detection signal: models exploiting the violation of cross-modal synchronisation patterns consistently outperform unimodal baselines~\cite{koutlis2024dimodif}. For AIGM, analogous inconsistencies may arise between AI-generated melody and lyrics when each is produced by separate generative processes, as is the case in systems underlying 
SONICS~\cite{rahman2024sonics}.

Among fusion strategies, early fusion~\cite{arevalo2017gatedmultimodalunitsinformation, kim2021stbertcrossmodallanguagemodel} is most appropriate when cross-modal mismatches are expected at the feature level, such as when AI-generated lyrics fail to align with the tonal key or rhythmic structure of the melody. However, it requires aligning audio and text into a common embedding space, which is non-trivial given the structural differences between symbolic music features and natural language. Late fusion, which processes each modality independently before merging decisions, avoids this alignment problem but risks missing precisely the cross-modal inconsistencies that multimodal detection is designed to exploit~\cite{frohmann2025double}. Intermediate fusion, exemplified by cross-attention mechanisms such as the Multimodal Transformer~\cite{tsai2019MULT}, offers the most promising balance: it allows domain-appropriate encoders for each modality while capturing cross-modal dependencies at multiple levels of abstraction, from low-level acoustic artifacts to higher-level semantic mismatches between lyrics and melodic structure.

Existing multimodal work in music provides empirical support for this direction. Studies on emotion and mood classification~\cite{chen2020multimodal, das2020multimodal} consistently demonstrate that melody and lyrics carry complementary rather than redundant information, with multimodal models outperforming single-modality counterparts. Crucially, Wang et al.~\cite{Wang2024} show that explicitly modelling musical structure, the distinction between verses and choruses, via hierarchical fusion further improves performance, suggesting that structural organisation, which is among the properties most difficult for AI to replicate convincingly, is precisely what multimodal models are best positioned to exploit. Frohmann et al.~\cite{frohmann2025double} provide the most direct evidence: their DE-detect system, combining transcribed lyrics and audio speech features via late fusion, outperforms existing lyrics-based detectors while remaining robust to audio perturbations, demonstrating the practical viability of multimodal AIGM detection. Nevertheless, their approach detects AI-generated \emph{lyrics} rather than targeting melody--lyrics cross-modal consistency as a detection signal, leaving a huge gap for future work.

\section{Beyond Binary Detection}
\label{4}

While existing AIGM detection methods reviewed in Section~\ref{2} predominantly address the task as a binary classification problem, which remains the primary focus of this survey, recent work explores detection scenarios beyond this binary framing. This section surveys these emerging efforts.

\subsection{Emerging Detection Formulations}

In practice, hybrid scenarios are increasingly common: human--AI co-creation, partially AI-generated accompaniment paired with human vocals, and extensive post-processing of AI-generated material: making a single global binary label often insufficient. To address this, several initiatives have introduced non-binary detection tasks. The Song Deepfake Detection Challenge 2025\footnote{\url{https://music-ir.org/mirex/wiki/2025:Song_Deepfake_Detection}} targets AI-generated accompaniment and vocals simultaneously, explicitly encouraging joint modelling of both components, a direct step toward detecting partially AIGM. On segment-level detection, the ADD 2023 challenge~\cite{add2023} introduced partially fake audio as a detection target, requiring localisation of AI-generated segments within human-produced content. These challenges represent the first concrete moves toward non-binary AIGM detection, though scenarios such as human--AI co-creation and post-processing or remixing of AI-generated material remain largely unaddressed. Related tasks such as source attribution and singing voice deepfake detection lie outside the primary scope of this survey but share methodological ground with AIGM detection; we mention them here briefly for context without details.

\subsection{Interpretability}

Interpretability has been identified as a critical requirement for AIGM detection: current systems produce binary verdicts without identifying which musical properties, harmonic irregularities, rhythmic inconsistencies, or timbral artifacts, triggered the decision, making it difficult to validate their reasoning or build trust in their outputs~\cite{li2024detectingmachinegeneratedmusicexplainability}. In particular, without interpretable outputs, it is hard to verify whether a detector is exploiting musically meaningful signals or generator-specific artifacts that will fail to generalise. The emerging consensus is that explainability must be defined in musically meaningful terms: it is insufficient to identify which input regions activated a decision; what is needed is identifying which specific musical properties triggered the verdict, requiring evaluation frameworks that assess not only detection accuracy but the musical validity of the explanations produced.

\subsection{Societal Implications}

AIGM detection also carries considerable societal implications. This is emphasised by 82\% of music creators expressing concern that AIGM may threaten their livelihoods~\cite{apramcos2024ai}, and critics arguing that AIGM lacks the emotional resonance and depth found in human compositions~\cite{Briot2021AI_Music_Survey}. Accordingly, the kind of personal expression, as, for instance, found in Beethoven's expressions of struggle, can currently not be replicated by generative models. Reliable deepfake detection is thus essential for safeguarding artistic integrity and ensuring fairness. At the same time, AIGM tools offer genuine creative potential, supporting music-based psychological interventions~\cite{bradt2021music}, personalised soundtracks~\cite{webster2023promise}, and AI-assisted composition~\cite{hadjeres2017deepbachsteerablemodelbach}, making the development of detection mechanisms not a rejection of AI in music, but a necessary condition for its responsible and balanced integration.

\section{Open Challenges and Future Directions}
\label{5}

With the limited scope of existing research on AIGM detection, several concrete and field-specific challenges define the research agenda ahead.

First, benchmark development must go beyond data collection. Existing datasets such as FakeMusicCaps and SONICS cover limited generative systems and genres, and no existing benchmark evaluates detection under realistic post-processing conditions. Future work should address: \textbf{How should AIGM detection benchmarks be designed to evaluate robustness under real-world conditions such as compression, pitch-shifting, and remixing?} and \textbf{How can partially AI-generated and human--AI co-created music be systematically annotated and benchmarked at the stem or segment level?} Concretely, future benchmarks could adopt leave-(k)-generator-out training splits to assess cross-generator generalisation, report performance under a standardised suite of post-processing conditions including MP3/AAC compression, pitch shifting, reverberation, and repeated lossy re-encoding to evaluate robustness beyond clean conditions, extend evaluation to segment-level F1 and stem-level binary classification accuracy for partially AI-generated music, and jointly report AUROC, EER, and stratified accuracy across genres and song-duration groups rather than relying on a single headline metric.

Second, the adaptation of audio deepfake detection to AIGM requires resolving a fundamental ambiguity: \textbf{How can detectors distinguish AI composition, the generation of melodic, harmonic, and structural content, from audio detection? Should we focus more on feature engineering sets or model design for grasping more transferable information for better generalisation?} Music, as an art form, incorporates numerous musicological restrictions and subjective qualities. Direct transfer learning or fine-tuning of existing models is insufficient for these requirements. Thus, there is a need for domain-specific methods tailored explicitly to music analysis.

Third, music-theoretic features identified in Section~\ref{3} remain largely unexploited while existing detectors in audio, text, and image domains often rely on superficial features. The key question is: \textbf{Which intrinsic musical properties, such as harmonic complexity, long-range motivic development, or rhythmic expressiveness, provide the most generalisable detection signals across generators, genres, and production pipelines? How could these be effectively detected?}

Future work could address this question through a feature-to-model validation pathway. Candidate properties could first be operationalised using symbolic and audio-derived descriptors, including harmonic complexity, cadence patterns, tonal stability, motivic recurrence, rhythmic complexity, and long-range structural organisation. For example, harmonic and cadential cues could be operationalised through chord progressions extracted from audio using chord-estimation tools such as Chordino~\cite{mauch2010approximate}, tonal-tension curves derived from Lerdahl-style tension models~\cite{lerdahl2007modeling}, and graph-based representations of chord progressions. Motivic recurrence and long-range form could be probed using structural features such as repetition, novelty, and self-similarity, together with phrase- or segment-level boundaries obtained from music-structure analysis tools such as MSAF~\cite{nieto2015msaf}. Rhythmic complexity could be quantified using established rhythmic descriptors and onset-density statistics, while timbral dynamics could be characterised using spectral-flux trajectories. Cross-modal musical--textual consistency could further be assessed through audio--text embedding similarity using models such as MuLan~\cite{huang2022mulan}, or through music-captioning models trained on resources such as LP-MusicCaps~\cite{doh2023lp}. Afterwards, these representations could then be incorporated into structure-aware detection architectures. Hierarchical Transformers operating over phrase- or segment-level representations could model long-range dependencies across musical sections. Graph neural networks over chord-progression graphs could explicitly model harmonic and cadential relationships encoded in the graph structure. Segment-level contrastive learning, inspired by approaches such as CLMR~\cite{spijkervet2021contrastive}, could be designed to encourage invariance to selected musically irrelevant perturbations while retaining sensitivity to musicologically meaningful structure. Multimodal cross-attention over melody, lyrics, and audio representations could provide a natural framework for testing whether cross-modal alignment provides additional detection signals. The resulting detectors could be evaluated across generators, genres, and production pipelines, with feature ablations and controls for low-level acoustic confounds to determine whether these music-theoretic cues provide transferable detection signals rather than dataset-specific proxies.

Fourth, multimodal detection remains an open problem. \textbf{How can cross-modal inconsistency between AI-generated melody and lyrics be operationalised as a detection signal}, given that no existing system explicitly targets this?

Fifth, \textbf{how will AIGM detection systems remain effective as generative models explicitly adapt to evade detection?} The adversarial dynamic between generation and detection represents a critical and underaddressed challenge for AIGM. This arms race is likely to intensify. As detection methods become more widely deployed, commercial AIGM platforms may deliberately or through iterative improvement produce outputs that are harder to detect, for instance, by incorporating more realistic long-range harmonic structure. This mirrors the GAN-style dynamic already observed between generation and detection in other domains~\cite{upenn2023detecting}, and suggests that static detectors trained on fixed datasets will degrade over time. Future detectors capable of identifying emotional and structural deficits in AIGM could also guide the refinement of generative models in an adversarial manner, driving advances in both detection and generation.

In conclusion, AIGM detection represents a crucial and underexplored research area, with substantial potential for further development. We have reviewed the progression from audio deepfake detection to AIGM detection, highlighting key challenges and emerging research domains. Despite the proliferation of general AI detection models, there remains a notable gap in models specifically designed to address the detection of AIGM, particularly from a musicological perspective. Given the increasing significance of AI-generated content in the music industry and beyond, we emphasize the need for more dedicated research and advancements in this field. Further efforts in this direction are essential to enhance the robustness and accuracy of detection models, ultimately contributing to more reliable and comprehensive AIGM detection systems.

\sloppy
\bibliographystyle{IEEEtran}
\bibliography{reference}

\end{document}